\documentclass[10pt,letterpaper]{article} \usepackage{opex3}

\usepackage{cite}
\usepackage{amsmath,amscd,amsthm,amsfonts,amssymb}

\begin{document}

\title{Far Field measurement in the focal plane of a lens  : a cautionary note}

\author{Pierre Suret  and St\'ephane Randoux}

\address{Laboratoire de Physique des Lasers, Atomes et
  Mol\'ecules, UMR CNRS 8523, Universit\'e~Lille~1, Sciences et
  Technologies, F-59655 Villeneuve d'Ascq, France  }

\email{Pierre.Suret@univ-lille1.fr} 



\begin{abstract*}

We study theoretically the accuracy of the method based on the Fourier property of lenses that is commonly  used for the far field measurement . We consider a simple optical setup in which the far-field intensity pattern of  a light beam passing through a Kerr medium is recorded by a CCD camera located in the back focal plane of a thin lens. 
Using Fresnel diffraction formula and numerical
computations, we investigate the influence of a slight longitudinal
mispositioning of the CCD camera.
 Considering a coherent gaussian beam, we show that a
tiny error in the position of the CCD camera can produce a narrowing
of the transverse pattern instead of the anticipated and
well-understood broadening. This phenomenon is robust enough to
persist for incoherent beams strongly modified by the presence of
noise.  The existence of this phenomenon has important consequences
for the design and the realization of experiments in the field of
optical wave turbulence in which equilibrium spectra reached by
incoherent waves can only be determined from a careful far-field
analysis. 
In particular, the unexpected narrowing of the far field may
be mistaken for the remarkable phenomenon of classical condensation of
waves.  
Finally, we show that the finite-size of optical components
used in experiments produces diffraction patterns having wings
decaying in a way comparable to the Rayleigh-Jeans distribution
reached by incoherent wave systems at thermodynamical equilibrium.
\end{abstract*}

\medskip

\bibliographystyle{osajnl}



\section{Introduction}
\label{sec:intro}

More than one century ago,  experiments  made by Abbe and
Porter provided the first spatial Fourier analysis of an object by observing the transmitted light in the focal plane of a lens\cite{Porter:06,Goodman:05}.
From these pioneering works,  Fourier optics theory has been widely
used to describe the propagation of light through lenses, to determine
the resolution of optical instruments and  to elaborate image
processing techniques\cite{Elias:52,Elias:53,Duffieux:83,
  Goodman:05}. Fourier optics is based on the Fourier transform
property of lenses :  the electric field in the back focal plane is
the Fraunhofer diffraction pattern  {\it i.e.} the Fourier transform
of the electric field profile in the front focal plane
\cite{Rhodes:53,Goodman:05}.

In the paraxial approximation, the Fraunhofer diffraction pattern
corresponds to the so-called far-field profile, {\it i.e. } the
intensity profile observed very far away from the source. In optical
experiments,  the far-field pattern may be measured without any
lens. For instance, the far field pattern at the output of optical
fibers can be simply measured with a single point detector rotating 
on a circle having its center that coincides with the output end of the fiber
\cite{Gambling:76,Hotate:79,Samson:85,Freude:85}. Another  approach
has been used in some nonlinear optical experiments in which  far
field patterns of laser fields modified by nonlinear  interactions
have been directly observed on screens or cameras placed far  from the
nonlinear medium. \cite{Callen:67,Nascimento:06,Wulle:93, Honda:97,Denz:98,
  Neill:05}.

However observation of the Fraunhofer pattern in the focal plane of a
thin lens provides a simple finite-distance method for the exact
measurement of the Fourier spectral power density. Placing amplitude
or phase masks in the focal plane of the lens, manipulation of the
transverse characteristics of light beams can be additionnaly
implemented \cite{Zernike:34,Zernike:55,Goodman:05}. \\

 Nowadays the measurement of the far-field is widely used in various areas
of optical resarch. In particular, measurements of the far field intensity pattern can be used to determine some linear and nonlinear properties of optical
materials.  For instance the transverse index profile of optical
fibers can  be determined from a far field analysis \cite{Gambling:76,Hotate:79,Freude:85}. 
The sign of Kerr coefficient of nonlinear materials can be determined
from simple vizualisations of the far-field pattern of a gaussian beam passing
through the nonlinear sample \cite{Lucchetti:09}.

Beyond those conceptually simple measurements of optical properties,
the far-field  pattern plays an important role in nonlinear optics
since it  represents a major tool to investigate complex
spatiotemporal phenomena.  In particular, (transverse) phase matching
conditions that deeply influence nonlinear interactions such as second
harmonic generation can be precisely studied from far-field
observations \cite{Wulle:93}. In this context,  the observation of the
far-field pattern is often considered as a complementary  tool of the
observation of the near-field pattern, especially regarding  lasers
systems \cite{Birky:69}.

In the field of nonlinear dynamics, the concept of far-field is widely
used in theoretical and experimental studies of transverse pattern
formation \cite{Cross93}. Over the last thirty years, far-field
observation  has been extensively used to investigate pattern
formation in nonlinear  optical systems based on photorefractive or
liquid crystals
\cite{Abraham:90,Arecchi:99,Westhoff:03,Denz:03,Arecchi:95,Ramazza:96,Ramazza:97,Honda:97,Agez:05,Agez:06,Gentilini:12}.
Let us emphasize that far-field determination is not only used for
static patterns but it can also be very useful to analyse complex
spatio-temporal behavior occurring in X-waves formation,
filamentation phenomena \cite{Kolesik:05,Faccio:08} or dispersive
shock waves\cite{Wan:07,Gentilini:12}.\\ 

Optical wave turbulence has recently emerged as a field in which the
experimental determination of the transverse far-field pattern is of
crucial importance. Wave turbulence (WT) can be defined as the
non-equilibrium  statistical dynamics of ensembles of
nonlinearly-interacting dispersive waves \cite{Zakharov}. The
archetype of wave turbulence is the random state of ocean surfaces but
it appears in various systems involving {\it e.g.} capillary waves, plasma waves, or elastic waves \cite{Newell:11,Miquel:11}.

Recently, nonlinear optical (in particular Kerr-like) systems have
turned out to be used for the investigation of wave turbulence
\cite{Picozzi07,Connaughton05,Dyachenko92}. Several experimental
studies of wave turbulence and wave thermalization of incoherent waves have 
been achieved in 1D systems based on optical fibers 
\cite{Pitois06,Barviau06, Barviau09,Suret10,Suret:11} and liquid 
crystals \cite{Bortolozzo09,Laurie:12}.

In Hamiltonian systems (described for instance by 2D or 3D Nonlinear
Schr$\ddot{o}$dinger equations) WT theory predicts the existence of a state of
thermodynamical equilibrium associated with a (kinetic) energy
equipartition among the Fourier components of the transverse
field. The thermodynamical equilibrium distribution reached by the wave system is 
named the Rayleigh-Jeans distribution. It is characterized by a lorentzian Fourier
spectrum decaying according to a power law $k^{-2}=|\bf{k}|^{-2}$ for high-frequency wavectors $\bf{k}$ observed
in the far-field \cite{Zakharov,Pitois06,Picozzi07}. Additionally, the
thermodynamical equilibrium of nonlinear incoherent waves may exhibit
a phenomenon of (classical) {\it wave condensation}  analog to the Bose-Einstein
condensation of atoms \cite{Connaughton05,Dyachenko92}. The main
signature of this condensation process is the accumulation of (quasi-) particles in
the fundamental Fourier component $\bf{k \sim 0}$. The observation of the 
thermodynamical equilibrium distribution and of the process of wave condensation
represents currently an experimental challenge in fundamental physics. 

Sun {\it et al.} have very recently studied the propagation of 2D
incoherent waves interacting in a Kerr-like photorefractive crystal
\cite{Sun:12}. In particular they report the observation of far-field
intensity patterns measured with a camera placed in the focal plane of
a lens. When the strength of the nonlinearity is increased, Sun {\it
  et al.} have observed accumulation of light around the transverse
wavevector of lowest value ($\bf{k \sim 0}$ and its transverse coordinates $k_x \sim 0$ and  $k_y \sim 0$) and they have interpreted
this observation as the phenomenon of optical wave
condensation. Recording the  far-field intensity pattern over four
decades, they report a high-dynamic measurement in which they evidence
a spectrum decaying over more than  three decades according to a power
law of $k_x^{-2}$.  The authors interpret the observation of this
power law as the Rayleigh-Jeans distribution, a signature of
thermodynamical equilibrium of the optical waves \cite{Sun:12}.\\

In those incoherent nonlinear optical experiments, proofs of the
existence of the phenomena of wave condensation and thermalization
can only be achieved from the measurement of the far-field transverse
intensity pattern. In this paper, we investigate questions related to
the existence of some experimental uncertainties that necessarily
arise in the measurement of the far-field intensity pattern in the
focal plane of a lens. We focus on the measurement of transverse
spectrum (wavevectors) of a gaussian beam passing through a Kerr-like
medium. We show in particular that the far-field pattern dramatically
depends on the position of the observation screen (or camera). Our
main results concern the observation of the far-field in a plane very
close to (but not exactly in) the focal plane of the lens. When the
third order nonlinearity increases, we show from numerical simulations
that the observed far-field can narrow whereas the true far-field
measured exactly in the focal plane broadens. This phenomenon can 
lead to serious misinterpretations of the observations made in optical
turbulence experiments.

In Sec. \ref{sec:section1},  we describe the typical 
experiments in which the far field of light passing through a thin Kerr
medium is observed nearby the focal plane of a thin lens. We recall the
usual and simple theoretical description of light propagation in the 
setup under consideration \cite{Goodman:05}.
In Sec. \ref{sec:section2}, we present the main result of this
paper. Using the Fresnel diffraction formula, we compute the 
far-field pattern of a coherent gaussian beam passing through a thin
Kerr-like medium. As it is well-known, the Fourier spectrum
monotonically broadens with concentric rings when the strength of the Kerr effect
increases\cite{Callen:67,Durbin:81,Dabby:70,Santamato:84,Luogen:05,Nascimento:06,Ramirez:10}. 
This is actually what is obtained from numerical simulations but observing 
the transverse pattern in a plane slighly shifted from the focal plane, 
we contrarily observe a narrowing of the transverse pattern instead of a broadening. 
In Sec. \ref{sec:section3}, we show that this
phenomenon persists for an initially incoherent beam. This means
that experiments on optical wave condensation are dramatically
sensitive to the alignement of the experimental setup.
In Sec. \ref{sec:section4}, we show that light diffraction by the edges 
of the nonlinear crystal can also lead to a misinterpretation of power 
laws experimentally observed in the Fourier spectrum.\\

\section{Position of the problem}
\label{sec:section1}

\begin{figure}[htbp]
  \centering \includegraphics[width=14cm]{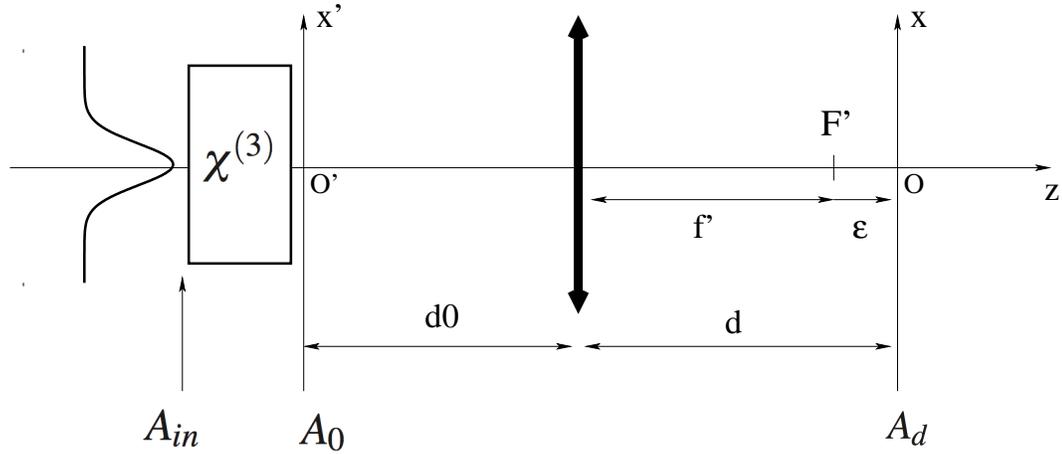}  
  \caption{\label{fig:setup}  \emph{Experimental setup.} The  incident
    field $A_{in}(x,y)$ passes through a thin Kerr ($\chi^{(3)}$)
    medium. $A_0(x',y')$ is the field in the output plane $(O'x'y')$
    of the $\chi^{(3)}$ medium. In order to measure the far-field of
    $A_0(x',y')$, the light propagates through a positive  thin lens
    with a focal length $f'$. $d_0$ is the distance between the plane
    $(O'x'y')$ and the lens. The transverse intensity profile
    $|A_d(x,y)|^2$ is observed in a plane close to the focal plane of
    the lens : $d=f'+\epsilon$ is the distance between the lens and
    the observation plane $(Oxy)$.  }
\end{figure}

In this article we consider a conceptually-simple setup found in many 
nonlinear optical experiments : the far-field of a light beam passing 
through a Kerr medium is measured in the focal plane of a thin lens
\cite{Arecchi:95,Ramazza:96,Ramazza:97,Sun:12}.  Fig. \ref{fig:setup}
represents a typical scheme of the setup under consideration.

The light beam propagating along the $z-$axis is supposed to be
monochromatic and  linearly-polarized. It is associated to an electric
field $E$ that reads:
\begin{equation}
\label{eq:E}
 E(x,y,z,t) = A(x,y,z) \, e^{j(k_0z-\omega_0t)}
\end{equation}
$A(x,y,z)$ is the complex amplitude of the field that is supposed to
be a slowly-varying  function of the longitudinal coordinate
$z$. $k_0=\omega_0/c=2 \pi / \lambda$ is the wavenumber, $c$ is the
speed of light in vacuum and  $\lambda$ is the wavelength of the
electric field in vacuum. In all the computations presented in this paper, 
 the numerical value taken for the wavelength $\lambda$ is $532$ nm.

The light beam passes through a transparent Kerr medium having a
length $L$.  In our discussion, the exact nature of this
Kerr medium has only little  importance : it can be either a piece of
bulk silica, a liquid crystal layer or a photorefractive crystal
\cite{Boyd:92}. The nonlinear effect under consideration can be either focusing or defocusing. The tranverse field incident onto the entrance side
of the  Kerr medium is $A(x,y,z=-L)=A_{in}(x,y)$. The transverse field
found at  the output side of the Kerr medium is
$A(x,y,z=0)=A_{0}(x',y')$ (see Fig. \ref{fig:setup}).

Let us consider that we want to record the transverse far-field
pattern  of the field $A_0(x,y)$ by using the standard and simple setup shown in Fig. \ref{fig:setup}. Let us recall that the
so-called {\it far field} $n(k_x,k_y)$  is the spectral power density
of $A_0(x,y)$. It is simply defined as 
\begin{equation}
n(k_x,k_y)=\Big |\widetilde{A_0}\big (k_x,k_y\big)\Big|^2
\end{equation}
where the transverse Fourier Transform (FT) of the amplitude
$A_0(x,y)$ is defined by:
\begin{equation}\label{eq:2Dfourier}
FT \Big ( A_0(x,y) \Big) = \widetilde{A_0} (k_x,k_y) =
\int\int_{-\infty}^{+\infty}A_0(x,y)\,e^{-i (k_x x+k_y y)}\,dx\,dy
\end{equation}

In order to measure the far-field pattern, the light beam passes
through a perfectly stigmatic thin lens having a focal length $f'$ and
adding a power-independent quadradic  term to the transverse phase of
$A_0(x,y)$. The intensity profile $|A_d(x,y)|^2$ is observed with a
camera in the plane $(O,x,y)$. As shown in Fig. \ref{fig:setup},
$d_0$ is the distance between the output side of the Kerr medium and
the thin lens wheras $d$ is the distance between the thin lens and the
camera.

Let us briefly recall why the transverse profile $|A_d(x,y)|^2$
recorded exactly in the focal plane of the thin lens ($d=f'$, $O=F'$)
coincides with the modulus square of the transverse Fourier transform
of $A_0(x,y)$. We will also examine how to calculate the transverse
intensity profile of the beam in a plane ``out of focus'',  {\it
  i.e. } shifted from the focal plane ($d=f'+\epsilon$).

Considering optical wave propagation under the paraxial approximation,
it can be easily shown (see \cite{Goodman:05} and Sec.
\ref{sec:annexe} for details)  that the transverse intensity profile
found in a plane separated by a distance $d$ from the lens reads : 
\begin{align}
\label{eq:nd}
\nonumber \big | A_d(x,y)\big |^2 = &\frac{1}{(\lambda d)^2} \\ \times
\;\Bigg | \;&\iint_{-\infty}^{+\infty}A_L(x',y')\,
\exp\Bigl[j(\frac{1}{d}-\frac{1}{f'})\frac{k_0}{2}(x'^2+y'^2)\Big]
\exp\bigl[-j\frac{k_0}{d}(xx'+yy')\bigr]\,dx'\,dy'\Bigg | ^2
\end{align}
where the FT of $A_L(x',y')$ is
$\tilde{A_L}(k_x,k_y)=\tilde{A_0}(k_x,k_y) \exp\bigl[-j
  d_0(k_x^2+k_y^2) / (2 k_0) \bigr]$.\\

If a CCD camera is placed exactly in the focal plane of the lens
($d=f', \epsilon=0$ in Eq. \ref{eq:nd}), it records an intensity profile $\big |
A_d(x,y) \big |^2$  that is proportional to the far field spectrum
previously defined :
\begin{equation}
\label{eq:spectrum}
n(k_x,k_y)=(\lambda f')^2\;  \Bigg| A_{f'} \Big(x=\frac{ f'}{k_0}k_x,
y = \frac{f'}{k_0}k_y \Big) \Bigg|^2
\end{equation}

If the CDD camera is not exactly placed in the focal plane of  the
lens ($d \simeq f'$ with $d \ne f'$), the term
$\exp\Bigl[j(\frac{1}{d}-\frac{1}{f'})\frac{k_0}{2}(x'^2+y'^2)\Big]$
found in Eq. (\ref{eq:nd}) plays a non-negligible role and the
intensity pattern recorded by the CCD camera can significantly
differ from the true far-field pattern recorded for $d=f'$.  As
described more in detail in Sec. \ref{sec:section2}, the  differences
between the transverse intensity pattern recorded  by the CCD camera
and the true far-field pattern increase  when the third-order
nonlinearity $\chi^{(3)}$ increases.

Before describing more in details the influence of a mispositioning of
the CCD camera, it is worth noticing that Eq. (\ref{eq:spectrum}) does
not depend on the distance $d_0$ between the output side of the
nonlinear medium and the lens when $d=f'$ ($\epsilon=0$).
Our numerical calculations have shown that the transverse intensity
profile recorded  out of the focal plane ($\epsilon \ne 0$)
depends slightly  on the distance $d_0$. Therefore, for the sake of simplicity,
we will not consider the influence of this additional parameter in
this paper and we will restrict our analysis to the situation  in
which the output side of the Kerr medium is close to the lens
($d_0=0$).

Passing through the Kerr medium, the beam acquires a nonlinear phase
depending on the transverse coordinates. In this paper, we limit our
theoretical work to the case of a thin nonlinear medium. Using the
common approach to describe propagation through thin nonlinear media
\cite{Castillo:94,Samad:98,Khoo:87,Sheik-bahae:89,Sheik-bahae:90,Deng:05,Ramirez:10},
\emph{we assume that light diffraction plays a negligible role in the
  nonlinear medium}. The length $L$ of the Kerr medium is thus
considered to be much smaller than the diffraction length. Under this
assumption, the complex amplitude of the light field found at the
output of the Kerr medium reads: 
\begin{equation}
\label{eq:kerrlens0}
  A_{0}(x,y)=A_{in}(x,y) \exp\Bigl[j \,2 \pi \, \gamma \,
    |A_{in}(x,y)|^2  \Big].
\end{equation}
$\gamma$ is given by $\gamma=\chi^{(3)}  L I_0 / (2 \pi)$ where
$\chi^{(3)}$ is   third-order nonlinearity coefficient. $|A_{in}|^2$
and $|A_0|^2$ are normalized with respect to the maximum value of the
optical intensity $I_0$. $\chi^{(3)}$ is positive in media such as
silica and it can be either positive or negative in photorefractive
media.

\section{Far field measurement  of a gaussian beam with self-phase modulation}
\label{sec:section2}

In this Sec. \ref{sec:section2} we examine the impact of a mispositioning of the  CCD
camera on the measurement of the far-field profile of a gaussian beam
having a waist $w$. The normalized amplitude of the electric field at
the input side of the nonlinear medium is :
\begin{equation}
\label{eq:gaussian}
A_{in}(x,y)  = \exp\big(- (1/2)\,(x^2+y^2)/w^2 \big )
\end{equation}
The numerical value taken for the diameter $2w$ of the gaussian beam
is $3$ mm (see Fig. \ref{fig:spectra}(a)), which is representative  of
the experiments on optical wave turbulence carried out by Sun {\it et al} 
in ref. \cite{Sun:12}.  We will examine the influence of
nonlinear phase shifts on the transverse structure of the light beam in 
a realistic way. In particular, in our numerical simulations, the maximum 
nonlinear phase shift taken by the light beam does not exceed $2 \pi$ 
($|\gamma|\le 1$), which corresponds to values readily accessible 
in standard experiments made with cw lasers passing
through liquid crystals or photorefractive crystals \cite{Arecchi:99,Sun:12}. In all this paper we use a numerical value of the focal length of the thin lens  $f'=50$mm.

All numerical calculations presented in Sec.
\ref{sec:section2}, \ref{sec:section3} and \ref{sec:section4} have been performed  
with transverse grids having a square shape. In Sec. \ref{sec:section2} and \ref{sec:section3} the surface of the square is $25$ mm$^2$  
($x,y \in [-5mm,5mm]$). All the numerical computations presented in this paper have
been performed with a large number of points ($32,768 \times 32,768$
points) in order to avoid any sampling and/or boundary conditions
artifacts (see Sec. \ref{sec:annexe}).

\begin{figure}[htbp]
  \centering \includegraphics[width=12cm]{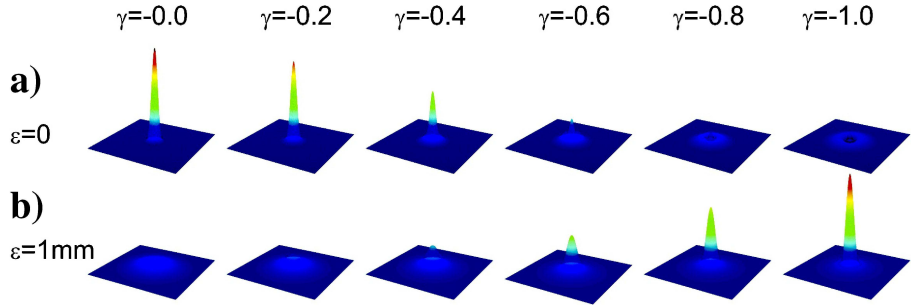}  
  \caption{\label{fig:2Dspectra} \emph{(a) Far-field patterns observed
      exactly in the focal plane of the lens ($\epsilon=0$) 
    and (b) transverse ``far-field'' patterns observed
      slightly out of the focal plane ($\epsilon=+1$mm).}
    The strength of nonlinearity is increased from $\gamma=0$ to
    $\gamma=-1$ (defocusing self-phase modulation). The far-field
    observed exactly in the focal plane of the lens broadens as the nonlinearity
    increases.  If the observation plane is slightly shifted ($\epsilon=+1$mm), 
    the observed ``far-field'' narrows when the nonlinearity increases.}
\end{figure}

First considering that the observation screen or CCD camera is well
positioned ($d=f'$),  we numerically compute the far field given by
Eqs.  (\ref{eq:spectrum}), (\ref{eq:kerrlens0}) and
(\ref{eq:gaussian}) for a lens having a focal length $f'=50$mm.  
Fig. 2(a) shows the evolution of the far field
pattern when the nonlinear  coefficient $\gamma$ is changed from zero
to $-1$ (defocusing self-phase modulation).  The light beam
experiences self-phase modulation and the far-field pattern thus
undergoes a significant broadening  when the nonlinearity increases.
As shown in Fig. 2(a), the far field pattern exhibits concentric rings
together with this spectral broadening. This phenomenon is very
well known and it has been extensively studied experimentaly and
theoretically  in the self-focusing and
self-defocusing cases \cite{Callen:67,Durbin:81,Dabby:70,
Santamato:84,Luogen:05,Deng:05,Nascimento:06}. Recently the influence
of a nonlocal Kerr effect on the far field of a gaussian beam has also been
considered \cite{Ramirez:10}

Let us now examine the consequences of a slight mispositioning  of the
observation screen (CCD camera) and let us assume that it is slightly
moved away from the lens,  at a distance $d=f'+\epsilon$ ($\epsilon >
0$) slightly larger than $f'$ (see Fig. \ref{fig:setup}). As shown in
Fig. 2(b), keeping $f'=50$mm, fixing $\epsilon=+1$mm and changing the nonlinear
coefficient $\gamma$ from zero to $-1$,  we now observe a narrowing of
the spectrum recorded by the CCD camera instead of the broadening
observed when the true far field is accurately measured (see
Fig. 2(a)).

\begin{figure}[htbp]
  \centering \includegraphics[width=12cm]{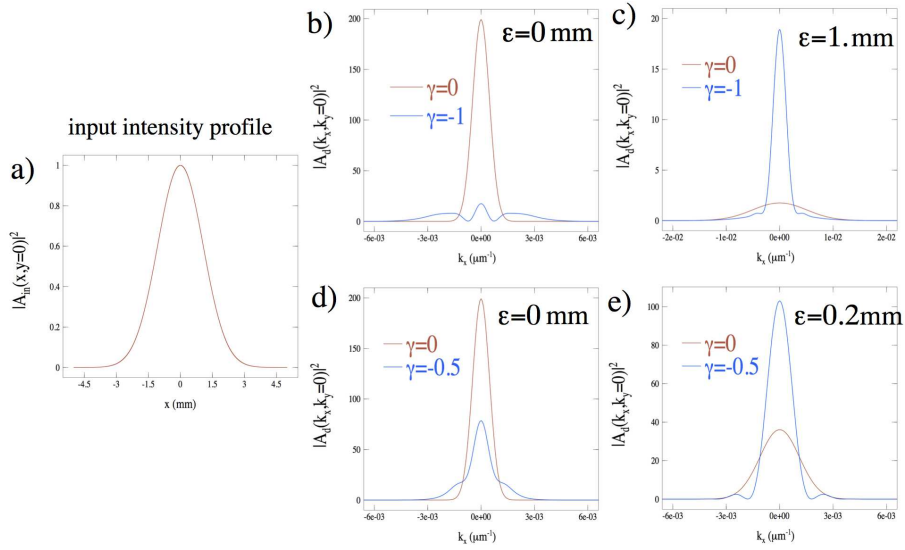}  
  \caption{\label{fig:spectra} \emph{Transverse intensity profiles}.
    a) Input profile.  b) and d) Far field observed exactly in the
    focal plane $\epsilon=0$ for a zero ($\gamma=0$, red curve) and
    nonzero third order  nonlinearity (blue curve b) $\gamma=-1$ d)
    $\gamma=-0.5$). c) and e) Corresponding transverse patterns
    observed sligtly out of the focal plane. c) $\epsilon=1$mm,
    $\gamma=0$ (red curve) and$\gamma=-1$ (blue curve). e)
    $\epsilon=0.2$mm, $\gamma=0$ (red curve) and$\gamma=-0.5$ (blue
    curve).}
\end{figure}

Moving from a three-dimensional plotting to a two-dimensional plotting, 
Fig. \ref{fig:spectra} shows transverse intensity profiles (cross sections with $y=0$ or $k_y=0$) for
some selected values of the mispositioning parameter $\epsilon$ and of the nonlinearity
coefficient $\gamma$. Fig. \ref{fig:spectra}(a) represents the intensity profile
$|A_0(x,y=0)|^2$ found at the output side of the nonlinear medium (near
field intensity pattern). The curves plotted in red lines in Fig. \ref{fig:spectra}(b), 
\ref{fig:spectra}(c), \ref{fig:spectra}(d), \ref{fig:spectra}(e) represent 
profiles recorded at several positions of the CCD camera without 
nonlinearity ($\gamma=0$). The curves plotted in blue lines in Fig. \ref{fig:spectra}(b), 
\ref{fig:spectra}(c), \ref{fig:spectra}(d), \ref{fig:spectra}(e) represent 
profiles recorded by the CCD camera at the same positions but for two 
possibles values of the Kerr coefficient ($\gamma=-0.5$, $\gamma=-1$).
As already evidenced in Fig. \ref{fig:2Dspectra}, the spectrum narrows instead of broadening
when $|\gamma|$ increases. This phenomenon is very sensitive and it can
be observed even for a very small shift of the camera out of the focal plane
($f'=50$ mm and $\epsilon=+1$ mm in Fig. \ref{fig:spectra}(c)). 
Fig. \ref{fig:spectra}(e)) shows that
an error $\epsilon$ in the position of the CCD camera as small as $\epsilon=+200
\mu$m leads to the same spurious observation of a spectral narrowing
instead of a spectral broadening.\\

It is possible to understand the phenomenon above reported from a
simple description made in terms of thin lenses combination. Let us 
firstly remark that, with our numerical parameters ($w=3$ mm, $\lambda=532$nm), the diffraction (Rayleigh) length is $\pi (2 w)^2 / \lambda \simeq 26$ m, which is
much greater than  any other length-scale involved in our problem.
If we first ignore optical Kerr effect ($\gamma=0$), this means 
that the input beam can be considered as being nearly collimated and consequently, 
that the position of the minimum beam waist is located very close to the 
back focal plane of the lens.

On the other hand, if $\gamma \ne 0$, self-phase
modulation of the gaussian beam inside the thin Kerr medium induces a
Kerr-lens \cite{Boyd:92} having a focal length $f'_{Kerr}$.
$f'_{Kerr}$ can be evaluated from a first-order Taylor expansion of Eq.
(\ref{eq:gaussian}) in $(x^2+y^2)/w^2$. Under this approximation, Eq. (\ref{eq:kerrlens0})
becomes~:
\begin{equation}
\label{eq:kerrlens0bis}
  A_{0}(x,y)=A_{in}(x,y) \exp (j \, 2\pi  \,\gamma \,  )\,
  \exp\Bigl[-j \, \gamma \,\pi \,(x^2+y^2)/w^2 \Big]
\end{equation}
The comparison between Eq. (\ref{eq:kerrlens0bis}) and Eq.
(\ref{eq:lens}) found in Sec. \ref{sec:annexe} gives :
\begin{equation}
\label{eq:fkerr}
f'_{Kerr}=\frac{w^2}{\gamma \lambda}.
\end{equation}
Still considering that $d_0=0$  (see Fig. \ref{fig:setup}), 
the gaussian beam passes through an effective lens made up 
with two lenses placed side by side with focal lengths 
respectively of $f'$ and $f'_{Kerr}$. The focal length of this effective lens is  given by
$(f'_{eff})^{-1}=(f')^{-1}+(f'_{Kerr})^{-1}$. In the examples
above presented (see Fig. \ref{fig:2Dspectra} and Fig. \ref{fig:spectra}),
the sign of the nonlinear coefficient is negative ($\gamma\le 0$) 
so that $f'_{Kerr} \le 0$ and thus $f'_{eff}>f'$. With
$f'=50$mm and $\gamma=-1$ one gets $f'_{Kerr}\simeq -4$m and
$f'_{eff}=50.6$mm.  Increasing the strength of the nonlinearity, the major 
consequence of the Kerr-lens effect is to push
the minimum beam waist at a distance $d=f'_{eff}>f'$ ({\it i.e.} $\epsilon
> 0$) from the lens.  Observing the transverse intensity pattern 
with a CCD camera whose position is fixed slightly out of the focal plane of 
the observation lens ($d=f'+\epsilon$, $\epsilon>0$), the transverse diameter of the pattern 
decreases when $|\gamma|$ increases, as illustrated in Fig. \ref{fig:2Dspectra} and
\ref{fig:spectra}.

Using numerical simulations, we have explored wide ranges of parameters and the unexpected phenomenon of
 narrowing of the far-field pattern observed slighlty out of the
focal plane of the thin lens has been found to be robust. In particular, the value taken by $d_0$
does not play any crucial role. Our numerical simulations also show
that the evolution of the measured pattern (with $\epsilon\ne 0$) is
non monotonic with $\gamma$. Increasing the value of $\gamma$ from
zero, the measured spectral width first decreases and then increases at
high value of $\gamma$. Our numerical simulations show that 
identical behaviors are also observed for  $\gamma>0$ and $\epsilon<0$. 

This can be easily understood from the 
interpretation above given and based on the effective focal length $f'_{eff}$.  
Note that if $\gamma$ and $\epsilon$ have identical sign, the width of the measured
spectrum increases as expected  but it strongly differs from the width of the true
far-field spectrum. From the general point of view, \emph{the error made in the measurement of the far field increases when the nonlinearity (self
phase modulation)  increases}.\\

In the phenomenon of wave condensation predicted from WT theory, 
the Fourier power spectrum (far-field) of an initially 
incoherent light wave narrows as the result of the turbulent nonlinear 
interactions among the waves \cite{Connaughton05,Dyachenko92}. 
The phenomenon demonstrated in this
paper has to be taken into account with great care in order to
interpret results from experiments designed to observe the 
process of optical wave condensation \cite{Sun:12}.

 In wave turbulence experiments tracking {\it e.g.} the wave condensation phenomenon, the initial condition is a incoherent {\it i.e.} random field \cite{Suret10,Suret:11,Sun:12}. In the Sec. \ref{sec:section3}, we explore the robustness of the unexpected narrowing of the measured far-field when strong phase noise is added to the gaussian field.

\section{Random initial condition}
\label{sec:section3}

In Sec. \ref{sec:section2}, we have considered the propagation of a coherent gaussian 
beam passing through a Kerr-like medium with an infinite transverse
size. We are now going to examine the nonlinear propagation of an initially incoherent
(i.e. noisy) optical beam in the same setup (see Fig. \ref{fig:setup}).   
The situation under consideration is in its principle very similar 
to the archetype of an experiment designed to observe the phenomenon of 
2D optical wave condensation\cite{Zakharov,Picozzi07,Connaughton05,Dyachenko92}. 
For instance, in experiments reported in \cite{Sun:12}, a gaussian beam passes through 
a spatial light modulator having a random transmission permitting to 
design an incoherent wave with an approriate Fourier spectrum. 

In this Sec. \ref{sec:section3}, we show that adding phase noise to the gaussian beam
does not qualitatively change the unexpected phenomenon of  far field narrowing decribed in Sec. \ref{sec:section2}.

\begin{figure}[htbp]
  \centering \includegraphics[width=12cm]{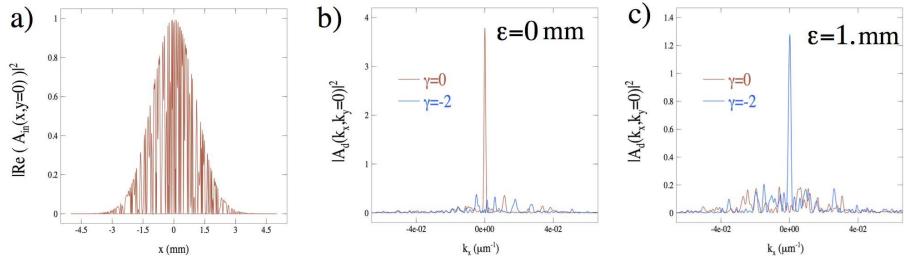}  
  \caption{\label{fig:noise} \emph{Random initial conditions} a)
    transverse intensity profile of the field at the entrance side of the Kerr medium 
    $|Re\;(A_{in}(x,y=0))|^2$ b)  Far field observed exactly in
    the focal plane ($\epsilon=0$) for $\gamma=0$ (red curve) and
    $\gamma=-2$ (blue curve) : the far-field broadens as the
    nonlinearity increases. c) Transverse patterns
    observed sligtly out of the focal plane $\epsilon=1$mm for $\gamma=0$
    (red curve) and $\gamma=-2$ (blue curve). Slightly out of the focal
    plane, the observed ``far-field'' narrows when the nonlinearity
    increases. This phenomenon is robustly preserved even if a strong phase noise
    is added to the initial condition. }
\end{figure}

To examine the nonlinear propagation of a spatially-incoherent beam, we replace 
the expression of the input gaussian amplitude given by Eq. (\ref{eq:gaussian}) by 
\begin{equation}
\label{eq:gaussiannoise}
A_{in}(x,y)  = \exp\big(- (1/2)\,(x^2+y^2)/w^2 \big ) \; \exp \big(j\;
\phi(x,y) \big )
\end{equation}

where $\phi(x,y)$ is a random phase. Fig. \ref{fig:noise}(a) represents the modulus square of the real part of $A_{in}(x,y=0)$ {\it i.e.} of the incoherent wave launched inside  the Kerr medium. 

Keeping mispositioning parameter  already used to compute Fig. \ref{fig:spectra}(b) and \ref{fig:spectra}(c) ($\epsilon=0$mm, $\epsilon=+1$mm) and
using $\gamma=-2$, we redo numerical simulations presented in Sec. \ref{sec:section2}
with this new initial condition. As in Sec. \ref{sec:section3}, we still use 
a transverse grid having a square shape ($x,y \in
[-5mm,5mm]$) discretized by $32768 \times 32768$ points. Fig. \ref{fig:noise}(b)
represents the far-field patterns observed exactly in the focal plane
of the lens with and without any nonlinearity (red and blue line
respectively). In spite of the presence of noise, Fig. \ref{fig:noise}(b) reveals 
features qualitatively similar to those already reported in Fig. 3.b for the coherent 
gaussian beam : the far-field intensity pattern broadens when the strength 
of the nonlinearity is increased. Moving the observation plane slightly 
out of focus ($\epsilon=+1$mm), Fig. \ref{fig:noise}(c) shows the evolution of the 
transverse intensity pattern when the strength of the nonlinearity is increased.  
Note that we have observed behaviors that are qualitatively similar by changing the phase noise into intensity noise. 

The presence of noise does not qualitatively change features 
already evidenced in Fig. 3(c) or 3(e): with a tiny error in the positionning of the detector,
the observed intensity pattern  narrows when $|\gamma|$ is increased.
This is a crucial result in the context of optical wave condensation experiments in which a speckle-like beam is launched inside a Kerr medium.

As an example, in some recents experiments made with a photorefractive
crystal, the strength of nonlinearity can be experimentally controlled
by the voltage applied to the crystal \cite{Sun:12}. The authors have 
observed a pronounced narrowing of the intensity pattern recorded by 
the CCD camera when the strength of the nonlinearity is increased. They interpret this spectrum narrowing as the classical condensation of waves arising from the interplay of nonlinearity and diffraction\cite{Connaughton05}. Neglecting diffraction inside a pure thin Kerr medium, our calculations show that this narrowing phenomenon can be observed in a plane very close to the focal plane of a thin lens.

\section{Finite-size  effect of optical components}
\label{sec:section4}

In Sec. \ref{sec:section2} and \ref{sec:section3}, we have considered 
the propagation of coherent and incoherent gaussian 
beams passing through a Kerr-like medium having an \emph{infinite} transverse
size.  In wave turbulence, Fourier spectrum is used in log scale and measured 
over many decades in order to observe power laws ; in particular, for
systems described by nonlinear Schr\"odinger  equations, the signature of wave thermalization 
is  a power law $k^{-2}$ for high transverse wavenumbers. We show in this Sec. \ref{sec:section4} that the \emph{finite} size  of the 
nonlinear medium itself can significantly influence the shape of the 
tails of the intensity pattern recorded by the CCD camera in the setup 
shown in Fig. \ref{fig:setup}. We then comment the way through which the 
observation of the resulting transverse intensity pattern can possibly 
lead to some misinterpretations in experiments on thermalization of optical waves.

\begin{figure}[htbp]
  \centering \includegraphics[width=12cm]{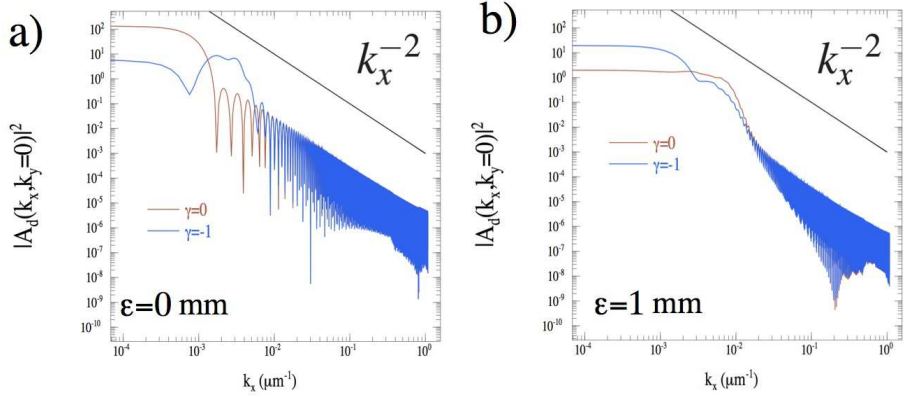}  
  \caption{\label{fig:fig5} \emph{Influence of the finite size of a square aperture.}
    An input gaussian beam passes through the kerr medium with a square
    section $a \times a = 5 \times 5$mm. Far field pattern observed
    (with log scale)~: a) exactly in the focal plane of the lens
    ($\epsilon=0$mm) b) slightly out of the focal plane
    ($\epsilon=1$mm). Red curves : $\gamma=0$. Blue curves :
    $\gamma=-1$. The phenomenon of unexpected narrowing of the spectrum
    is again observed for $\epsilon=1$mm. One observes the signature
    of the Fourier tranform of the square
    $|\sin(k_xa/2)/(k_xa/2)|^2$. Black curves : $1/k_x^2$}
\end{figure}

We assume that the nonlinear bulk medium  has a square section with $5$mm-long
sides. As in Sec. \ref{sec:section2}, we consider the nonlinear propagation of a coherent gaussian beam. 
The extension of the beam transverse profile is spatially limited and the input
amplitude of the field $A_{in}(x,y)$ described by equation
(\ref{eq:gaussian}) is now changed into:
\begin{eqnarray}
\label{eq:gaussian2}
&A_{in,square}(x,y)  &=  rect(x,y) \times A_{in}(x,y) \\
\label{eq:rect}
\text{with} \;\;\;\;\;\;\;\;\;\; & rect (x,y) &= 
\begin{array}{l}
  1  \;\;\;\;\;\;\;\text{if}\; |x|\le a\;\text{and}\; |y|\le a
  \;\;\;\; \text{with} \;\;\; 2a=5 \text{mm} \\ 0
  \;\;\;\;\;\;\;\text{else}
\end{array}
\end{eqnarray}
Keeping parameters already used to compute Fig. 3(b) and 3(c) ($\epsilon=0$mm, $\epsilon=+1$mm and
$\gamma=-1$), we redo numerical simulations presented in Sec. \ref{sec:section2}
with this new initial condition. As in Sec. \ref{sec:section2}, we still use 
a transverse grid having a square shape (but with $x,y \in
[-7.5mm,7.5mm]$) discretized by $32768 \times 32768$ points. 
Fig. 5(a) plotted in vertical logarithmic scale represents far-field patterns observed
exactly in the focal plane of the lens ($\epsilon=0$mm) with and without nonlinearity
(red and blue lines respectively).  Fig. 5(b) represents 
the transverse intensity pattern observed slightly out of focus ($\epsilon=+1$mm). As
already evidenced in Sec. \ref{sec:section2}, Fig. 5(b) shows that the spectrum narrows instead
of broadening when the strength of the nonlinearity increases.

However let us emphasize that the wings of the 
transverse intensity spectra decay according to a power law scaling as $k_x^{-2}$
whatever the position of the CCD camera. This new feature arises from the 
diffraction of light onto the edges of the finite-size nonlinear medium. 
The origin of the power law observed and of the exponent $-2$ can be easily 
understood from simple physical considerations. The Fourier transform 
of $A_{in,square}(x,y)$ is the convolution of the Fourier transform of $A_{in}(x,y)$ 
with the Fourier transform of $rect(x,y)$ which is $\widetilde{rect}(k_x,k_y,)=(2a)^2 \,sinc (k_xa)\, sinc (k_ya)$. Reminding that  $sinc(x)=\sin(x)/x$, the maxima of the horizontal cross section of the power spectrum are propotionnal  to $1/k_x^2$.

Although this decay of the wings of the spectra according to a power law 
can be straigthforwardly interpreted from light diffraction onto the edges 
of the nonlinear medium, we emphasize that this behavior must be 
well kept in mind in optical wave turbulence experiments. Indeed a decay
of a power spectrum according to a $1/k^2$ law is the
signature of the thermodynamical Rayleigh-Jeans distribution characterizing 
the equilibrium state reached by wave
systems described by NLS equations \cite{Zakharov,Picozzi07,Sun:12}.

Note that  Rayleigh-Jeans distribution predicted by WT theory applied to the 2D nonlinear Schr\"odinger equation corresponds to a power law  $k^{-2}=|\bf{k}|^{-2}$ for high-frequency wavectors $\bf{k}$ \cite{Zakharov,Picozzi07}. For the sake of simplicity, an horizontal cross section of the spectrum is plotted 
in Fig. \ref{fig:fig5}  as it seems to be done in \cite{Sun:12}. The horizontal axis is therefore the $k_x$ component of the wavevector ${\bf k}$. Averaging all the cross sections over the isoline $k=\sqrt{k_x^2+k_y^2}$ in the  ${\bf k}$-space gives a slightly modified power law for the spectrum $n(k)$. The main goal of this Sec.  \ref{sec:section4} is not an exaustive study of this power law but to point out that diffraction effect may play crucial role in 2D WT experiments.

In all optical experiments, light diffraction onto the edges of crystals, 
mirrors or lenses may lead to this kind of decay obeying a power law.
Therefore, wave turbulence experiments have to be designed with
caution : the spectrum tails corresponding to the diffraction on all the edges
found in the setup have to have a negligeable weight relatively to awaited
thermodynamical distribution.

\section{Conclusion}\label{sec:conclusion}

In this paper, we have theoretically studied a conceptually-simple 
optical setup in which the far-field intensity pattern of 
a light beam passing through a Kerr medium is measured in the focal plane of a thin lens. Using Fresnel diffraction formula and usual theoretical models of thin
lens and thin Kerr medium, we have investigated the influence of the nonlinear phase shift and of the distance between the lens and the detector.

We have shown that the transverse intensity pattern observed in the 
back focal plane of the thin lens dramatically depends on the longitudinal position 
the CCD camera. Although the ``true'' far field pattern is actually correctly measured 
when the CCD camera is exactly placed in the focal plane, significant deviations 
from the ``true'' far field pattern are observed even for a small mispositioning 
of the CCD camera. Those deviations have been shown to possibly occur 
in real-life optical systems, for light beams experiencing nonlinear phase shifts 
of the order of $2 \pi$, and for errors in the position of the CCD camera 
as small as $\sim 200 \mu$m or  $\sim 1$ mm at focal lengths of $\sim 50$ mm. From the general
point of view, the phenomenon reported in this paper may be at the origin of large uncertainties in the quantitative measurement of transverse wavevectors and it may take some
importance in various experiments devoted for instance to the study of
pattern formation \cite{Arecchi:99,Denz:03} or dispersive shock waves
\cite{Wan:07,Gentilini:12}.

However, far beyond the existence of quantitative errors possibly made in some measurements, 
we have shown that a tiny mistake in the position of the CCD camera
may lead to qualitatively-erroneous interpretations about the observed 
phenomena. Mispositioning the CCD camera by only $\epsilon\simeq 200 \mu$m
behind a lens with a focal length of $50$ mm and increasing the nonlinear 
phase shift from $0$ to $\sim 2 \pi$, numerical simulations evidence a 
narrowing instead of the anticipated broadening of the transverse intensity 
pattern. This phenomenon has been demonstrated for a coherent gaussian beam (see Sec.
\ref{sec:section2}) but it is robust enough to persist for incoherent 
beams strongly modified by the presence of a phase noise (see Sec.
\ref{sec:section3}). We believe that this result is of significance in
the context of optical experiments tracking the phenomenon of wave condensation
\cite{Connaughton05,Dyachenko92,Sun:12}. Optical wave condensation is
a classical phenomenon analogous to the bose Einstein condensation : its
signature in 2D optical experiments is a narrowing of the Fourier (far-field) 
spectrum as  the result of the interplay between diffraction and nonlinearity
in the Kerr-like medium.

In addition with these problems related to the positioning of the CCD camera behind the lens, we have shown that WT experiments may suffer from another artefact. Wave condensation and energy equipartion among the modes (with high wavenumbers $k$) are two phenomena arising from the same general wave thermalization process \cite{Connaughton05}. In 2D waves systems that are described by nonlinear Schr\"odinger equation, energy equipartition (characterizing the so-called Rayleigh-Jeans spectrum) corresponds to a power law $k^{-2}$ in the far field pattern \cite{Zakharov,Picozzi07,Connaughton05}. We have shown that the well-known  transverse intensity patterns arising from light diffraction by the edges of  the nonlinear medium may be wrongly interpreted as this Rayleigh-Jeans spectrum  (see Sec. \ref{sec:section4}).

Recent works have
shown that nonlinear optics is a field in which 
fundamental phenomena predicted by WT
theory  can be explored \cite{Picozzi07,Suret:11,Suret10,Pitois06,Sun:12}. 
Phenomena such as wave thermalization, wave condensation or Kolmogorov-Zakharov 
cascades have clear signatures in the Fourier space\cite{Zakharov}. In spatial optical experiments, 
the observation of the spectra of incoherent waves requires a far-field analysis. 
We have shown in this article that the observation of these
phenomena in transverse nonlinear optics experiments is very challenging ant that 
it requires great cares in the measurement of the far field spectrum. 

Let us emphasize that those difficulties in the measurement of the transverse far field pattern 
are usually not found in optical experiments dealing with transverses patterns 
such as rolls or hexagons which have usually a finite number of transverse 
wavevectors \cite{Cross93,Arecchi:99}. As a consequence, in the studies devoted to
pattern formation, experimentalist may sligtly adjust the position
of the detector in order to observe sharp and discrete components in
the far field. In the context of wave turbulence, the wave spectra are 
continuous and it is therefore hard to apply simple ``eye-criteria''
to align the setup.

Future experimental works will have to be done in order to observe the
unexpected narrowing of the far field numerically predicted in this paper. Moreover,
in future wave turbulence experiments in nonlinear optics, it is very important to go
through the limitations described in this paper. In particular, highly
sensitive methods have to be developped and used in order to adjust
the relative position of the lens and of the detector. Transverse analysis 
of the phase and not only of the intensity of the beam could be helpful 
to align the optical setup. \\

\section{Appendix: Fresnel diffraction and propagation through a thin lens}\label{sec:annexe}

In this Sec. \ref{sec:annexe}, we give a brief demonstration of Eq. (\ref{eq:nd}).
 Eq. (\ref{eq:nd}) gives the transverse intensity profile of a
linearly polarized field in a plane located at a distance $d$ from a
thin lens with a focal length $f'$ (see Fig. \ref{fig:setup}). We 
report the reader to \cite{Goodman:05} for a detailed demonstration.\\

In all this paper, we assume that the amplitude  $A(x,y,z)$ is a slowly-varying
function of the longitudinal coordinate ({\it i.e.} $k_0 |\partial_z A | >> |\partial_{zz} A
|$). Under this paraxial assumption, the usual propagation equation
for a linearly-polarized electric field in vacuum $\Big
(\partial_{tt}-c^2\partial_{zz} \Big ) E = 0$ becomes:

\begin{equation}
\label{eq:paraxial}
 \partial_z A = \frac{j}{2k_0} (\partial_{x}^2 + \partial_{y}^2) A
\end{equation}

From Eq.(\ref{eq:paraxial}), it is straightforward to show that the transverse Fourier transform of $A(x,y,z)$ reads:
\begin{equation}
\label{eq:Az}
 \widetilde{A}(k_x,k_y,z) = \exp \Big [ - \frac{j
     (k_x^2+k_y^2)}{2k_0}z \Big ] \;\;  \widetilde{A}(k_x,k_y,z=0)
\end{equation}
where $\widetilde{A}(k_x,k_y,z=0)$ is the transverse Fourier transform of $A(x,y,z=0)$.

Using the inverse Fourier transform, we get the so-called
\emph{Fresnel diffraction} formula :

\begin{equation}
\label{eq:fresnel}
A(x,y,z)=\frac{1}{j\lambda z}
\iint_{-\infty}^{+\infty}A(x',y',z'=0)\,\exp \bigl[j\frac{\pi}{\lambda
    z}((x-x')^2+(y-y')^2)\bigr]\,dx'\,dy'
\end{equation}

Propagating through a thin positive lens
used in Gauss conditions ({\it i.e.} in paraxial approximation), let us recall that the electric field acquires 
a quadratic phase \cite{Goodman:05} so that:
\begin{equation}
\label{eq:lens}
  A'_L(x,y)=A_L(x,y) \exp\Bigl[-j\frac{\pi}{\lambda f'}(x^2+y^2)\Big]
\end{equation}
where $A_L$ is the input field
in the front plane of the lens, $A'_L$ is the field just after the
the lens and $f'$ is the focal length of the lens, .\\

As considered in Fig. \ref{fig:setup}, the field $A(x,y,z)$
first propagates over a distance $d_0$, then pass through a lens with a
focal lens $f'$ and finally propagates over a distance $d$ (before reaching the
screen or the camera). Combining Eq. (\ref{eq:fresnel}) and (\ref{eq:lens}) one gets :
\begin{align}
\label{eq:Ad2}
\nonumber A_d(x,y)= &\frac{1}{j\lambda d}
\exp\Bigl[j\frac{\pi}{\lambda d}(x^2+y^2)\Bigr] \\ \times
\;&\iint_{-\infty}^{+\infty}A_L(x',y')\,
\exp\Bigl[j(\frac{1}{d}-\frac{1}{f'})\frac{\pi}{\lambda}(x'^2+y'^2)\Big]
\exp\bigl[-j\frac{2\pi}{\lambda d}(xx'+yy')\bigr]\,dx'\,dy'
\end{align}
If $d=f'$, $A_d(x,y)$ is proportionnal to the Fourier
transform of $A_L$ and finally one gets :
\begin{equation}
\label{eq:farfield1}
  |A_f'(x,y)|^2= \frac{1}{(\lambda f')^2} | \tilde{A_L}(k_x,k_y) |^2
  \; \text{with spatial wavenumbers} \;\; k_x=\frac{2\pi x}{\lambda
    f'} \;\; \text{and} \;\; k_y=\frac{2\pi y}{\lambda f'}
\end{equation}
where $\widetilde{A_L}(f_x,f_y)$ is given by Eq.
(\ref{eq:Az}) with $z=d_0$, $A(x,y,z=0)=A_0(x,y)$ and
$A(x,y,z=d_0)=A_L(x,y)$. Using $| \tilde{A_L}(k_x,k_y) |^2= | \tilde{A_0}(k_x,k_y) |^2$  and considering the field exactly  in the focal plane of the lens $A_{d=f'}(x,y)$ one finally gets :

\begin{equation}
\label{eq:farfield2}
  |A_{f'}(x,y)|^2= \frac{1}{(\lambda f')^2} | \tilde{A_0}(k_x,k_y) |^2
\end{equation}

Notice that the intensity far field pattern $|A_{f'}(x,y)|^2$ does not
depend on the propagation distance  $d_0$ between the initial plane
$(O',x',y')$ and the lens. $d_0$ only affects the transverse phase of
$A_{f'}(x,y)$.\\

In the numerical simulations, the propagation in free space is computed in the
Fourier space with the FFT routine FFTW \cite{fftw:05} and the
equation (\ref{eq:Az}). On the contrary, we compute the field passing
through a lens in the direct space from Eq.
(\ref{eq:lens}). The boundary conditions are periodic on our square
grid.
As suggested by equation (\ref{eq:farfield2}) and as it would be done in experiments, we finally multiply the intensity pattern $|A_d(x,y)|^2$ by the constant $(\lambda f')^2$ in order to get the spectral power density in the Fourier space.

Along the propagation, the beam experiences strong changes in its
typical diameter. The beam diameter varies from a few millimeters in the input plane ($w=3$mm) to
$10\mu$m in the focal plane of the lens. In order to avoid
sampling problem we have used a large number of points with a grid of
$32768 \times 32768$  points. We have carefully checked the accuracy of numerical computations (in particular those represented in log scale
in Fig. \ref{fig:fig5}) by varying the number of points and the  size of the window ($5\times5$mm in Sec. \ref{sec:section2}, \ref{sec:section3} and \ref{sec:section4}
and $7.5\times 7.5$mm in Sec. \ref{sec:section4}).

Finally, notice that when the circular symmetry is preserved, it is possible to simplify the equation (\ref{eq:Ad2}). $A_L(x',y')$ is replaced by $A_L(r)$ where $r=\sqrt{x'^2+y'^2}$. Using Bessel functions properties we get :

\begin{equation}
\label{eq:Bessel1D}
|A_d(x,y=0)|^2= \frac{1}{(\lambda d)^2}\;\Big |\int_{-\infty}^{+\infty}A_L(r)\,\times\,
\exp\Bigl[j(\frac{1}{d}-\frac{1}{f'})\frac{\pi}{\lambda}\,r^2\Big] \,\times\,
J_0 \Bigl( \frac{2\pi r x} {\lambda d}\Bigr )\,\times\,2 \pi \, r \, dr\Big |^2
\end{equation}

where $J_0(X)$ is the first kind modified Bessel function of order $0$. This formula can be applied only to examples of Sec. \ref{sec:section2} (circular symmetry) but not to the cases of Sec.  \ref{sec:section3} and  \ref{sec:section4} which require 2D calculations.

\medskip

{\bf Acknowledgment}

\medskip

The authors thank Serge Bielawski for fruitful discussions and computing facilities and Elise Berrier for the nice 3D picture. The authors are grateful to Guy Millot for fruitful discussions. The Laboratoire de Physique des Lasers Atomes et Mol\'ecules is Unit\'e Mixte de Recherceh 8523 of th CNRS. This work was supported by the Centre Europ\'een pour les Math\'ematiques, la Physique et leurs Interactions (CEMPI).

\end{document}